\documentclass[12pt]{iopart}
\usepackage{hyperref}
\usepackage{amssymb}
\begin{document}

\def\a{a}
\def\ak{a^\dagger}
\def\bra#1{\langle #1|}
\def\ket#1{| #1\rangle}
\def\skal#1#2{\langle #1| #2\rangle}

\def\bran#1{\langle\widetilde{ #1|}}
\def\ketn#1{\widetilde{| #1\rangle}}

\def\norma#1{e^{-{{{\vert #1 \vert} \over 2}}^2}}
\def\normaa#1{e^{{{{\vert #1 \vert} \over 2}}^2}}
\def\tay#1#2#3{{#1^{\!^{(\!#2\!)}}\!\!(#3)}}
\def\tayf#1#2#3{{{{#1^{\!^{(\!#2\!)}}\!\!(#3)}}\over{#2!}}}
\def\pocjed#1{\begin{equation}\label{eq:#1}}
\def\jed#1{(\ref{eq:#1})}
\def\op#1{{#1}}

\title{Construction of the phase operator using logarithm of the annihilation operator}
\author{Aleksandar Petrovi\'c}
\address{University of Belgrade, Serbia.}
\ead{a.petrovic.phys@gmail.com}

\begin{abstract}
We investigate a lema that excludes existence of the phase operator and present a condition to avoid the lema. A
method for construction of an analytic function $f$ of the annihilation operator $\a$ is given. $f(z)$ is analytic
on some compact domain that does not separate the complex plane. Using these results we obtain $\ln\op a$. Since
$\lbrack\op{a}^\dagger \op{a}\ ,-i\ln\op{a}\rbrack=i$, we can use $\ln\op a$ to construct an operator $\op{\Phi}$,
which satisfies the definition of the phase operator.
\end{abstract}

\section{Introduction}
The annihilation operator in Hilbert space is operator $\a$ which satisfies relation:

\pocjed{komutator}
\lbrack\a,\ak\rbrack=1.
\end{equation}
A self-adjoint number operator $N$ and the corresponding Fock states are constructed using $\a$:
\pocjed{operator_br_cestica} N=\ak\a, \ \ \ \ \  N\ket n =n\ket n ,\ \ \ n=0,1,2,3,...
\end{equation}
The phase operator is an operator conjugated to number operator:

\pocjed{def4}
\lbrack\op{N},\op{\Phi}\rbrack=i I.
\end{equation}
It is an operator which corresponds to the phase of the harmonic oscillator. Its eigenvalues are phases of the
quantum harmonic oscillator. An explicit form of this operator has not been given, and some authors have
questioned its existence \cite{carruthers},\cite{leonhardt}. 
Best approach to construction of the phase operator is given by Pegg and Barnett \cite{Pegg}.
They constructed an operator and corresponding eigenfunctions which give a good approximation in
finite dimensional space. However, when dimension of space tends to $\infty$, this operator does not exists as an operator on a Hilbert space.

\noindent 
Although form of the phase operator has not been given, basis of its
eigenvalues is well known \cite{levi}:

\pocjed{vektor_faze_def}
\ket{\varphi}_f = {1\over{\sqrt{2\pi}}}\sum\limits_{k=0}^\infty e^{ik\varphi} \ket{n}\ .
\end{equation}
The vectors $\ket\varphi_f$ form a non-orthogonal basis:

\pocjed{scal_proiz}
_f\skal{\varphi^\prime} \varphi _f={1\over{2}}\delta(\varphi^\prime - \varphi) +
{1\over{4\pi}}(1+i\cot{{1\over{2}}(\varphi^\prime - \varphi)}).
\end{equation}

\noindent
The $\ket\varphi_f$ vectors have a useful
property, which in fact recommends them as eigenvectors of the phase operator. Their time evolution is natural:

\pocjed{evolucija}
\e^{iNt}\ket\varphi_f=\ket{\varphi+t}_f .
\end{equation}
A recent result, giving an expression for analytic functions of the annihilation operator \cite{rad1_arxiv}, is
used here as a starting point for construction of the phase operator. Hence, for the first time we can construct a
correct expression for $\ln\op{a}$, which will then be used as a base for the further construction of the phase
operator.
\noindent Namely, an explicit construction of an operator which satisfies commutation relation
\jed{def4} and has vectors \jed{vektor_faze_def} as eigenvectors, is given.
\noindent
In section 2, we give a set
of conditions for operator $\op{\Phi}$ which circumvent the non-existence arguments. In section 3, we construct an
expression for analytic functions of the annihilation operator and discuss some of its properties. In section 4 we
construct $\ln\op{a}$. In section 5, starting from $\ln\op{a}$, we construct the phase operator $\op{\Phi}$.

\section{Existence of the phase operator}

The usual proof of non existence of the phase operator is based on reduction to contradiction \cite{leonhardt}.
Namely, assuming this operator exists, using \jed{operator_br_cestica} and \jed{def4} we obtain:

\pocjed{faza_n}
\bra{n}\lbrack\op{N},\op{\Phi}\rbrack \ket{m}=(n-m)\bra{n} \op{\Phi} \ket{m} \not = i I.
\end{equation}
\noindent However, if it is assumed that the vectors  $\ket{n}$ do
not belong to the domain of the operator $\op\Phi$, then argument
\jed{faza_n} breaks down. We see that a necessary condition for
existence of the phase operator $\op{\Phi}$ is:

\pocjed{kontra4}
\ket{n}\not\in D(\op{\Phi}),\ \ \ n=0,1,2,3,...
\end{equation}
leading to that $\op{\Phi}$ in $\ket n$ representation cannot be expressed as a matrix. If $\op{\Phi}$ exists, it
can be expressed as a product of two or more matrices. Indeed, in section 5, we will construct an operator
$\op{\Phi}$ that satisfies these conditions.

\section{Analytic functions of the annihilation operator}

In this section, we briefly describe a construction of function $f$ of the annihilation operator $\a$. $f(z)$ is
analytic on some compact domain that does not separate the complex plane. A more detailed analysis of this topic
is given in \cite{rad1_arxiv}. First, we give note about Runge's approximation theorem and a sequence of
polynomials which approximate $f(z)$ on the whole domain. Then we construct a new form of identity which is well
suited for construction of $f(\a)$. We proceed constructing $f(\a)$ and finish with an analysis of some of its
properties.

\subsection{Runge's approximation theorem}

{\bf Theorem:}  If $f$ is an analytic function on a compact domain $\Omega$ that does not separate the complex
plane, then there exists a sequence $P_l(z)$ of polynomials such that  converges uniformly to $f(z)$ on $\Omega$
\cite{saff},\cite{markushevic},\cite{alfors}:

\pocjed{runge1}
f(z)=\sum\limits_{l=0}^\infty
P_l(z-z_0)=\sum\limits_{l=0}^\infty\sum\limits_{k=0}^{d_l} c_k^{(l)}  (z-z_0)^k , c_k^{(l)}\in\mathbb{C}, z,z_0
\in \Omega.
\end{equation}
\noindent Runge's theorem  is an existence theorem, i.e. it does not give values of $c_k^{(l)}$. Functions which
can be approximated using polynomials \jed{runge1} are also $\ln z$ and $z^\lambda, \lambda\in\mathbb{R}$. In this
paper Mittag-Leffler expansion \cite{markushevic} is used to approximate function $\ln z$.

\subsection{New identity resolution}

Eigenstates $\ket\alpha$ of annihilation operator $\a$ are called coherent states:

\pocjed{koherent1}
\a\ket{\alpha} =\alpha \ket{ \alpha }  , \alpha\in\mathbb{C}.
\end{equation}
$\ket\alpha$ can be expressed in terms of Fock states $\ket{n}$:

\pocjed{koherent_preko_n}
\ket{\alpha}=\norma
\alpha\sum\limits_{n=0}^\infty {{\alpha^n}\over{\sqrt{n!}}} \ket n , \alpha\in\mathbb{C}.
\end{equation}
Non-normalized coherent states are:

\pocjed{nenormirana_stanja}
\ketn \alpha=\rme^{\alpha\ak} \ket 0 = \normaa \alpha \ket \alpha.
\end{equation}
Coherent states form an overcomplete and non-orthogonal set which spans the resolution of identity \cite{arsa}:

\pocjed{stara_jedinica}
I={1\over\pi}\int\rmd^2\alpha\ket\alpha\bra\alpha .
\end{equation}
In formal analogy with the spectral theorem one can write an entire function of the annihilation operator:

\pocjed{entire_function}
f(\a)={1\over\pi}\int\rmd^2\alpha f(\alpha)\ket\alpha\bra\alpha .
\end{equation}
However, expression \jed{entire_function} is not valid for non-entire functions \cite{arsa}, so we construct a new
identity resolution:

\pocjed{nova_jedinica1} I=-i\oint\displaylimits_{\vert\gamma\vert=R}
{{\rmd\gamma}\over{\gamma}}\ketn{\gamma+z_0}\bran\gamma\; J\;\rme^{-z_0 \ak}\ , R>0, z_0\in\mathbb{C},
\end{equation}
where

\pocjed{nova_jedinica2} J={1\over{2\pi}}\sum\limits_{n=0}^\infty{{n!}}\ket n\bra n  .
\end{equation}

\subsection{Construction of analytic functions of the annihilation operator}

Using sum \jed{runge1} and resolution of identity \jed{nova_jedinica1}, an analytic function of annihilation
operator can be constructed:

\begin{eqnarray}\label{eq:Mittag2}
f(\a) &=& \sum\limits_{l=0}^\infty\sum\limits_{k=0}^{d_l}
c_k^{(l)}
(\a-z_0)^k \cdot I\nonumber\\
&=& -i\sum\limits_{l=0}^\infty\sum\limits_{k=0}^{d_l} c_k^{(l)} \oint\displaylimits_{\vert \gamma\vert=R}\!\!\!
\rmd\gamma \gamma^{k-1} \ketn{\gamma+z_0} \bran\gamma \; J\; \rme^{-z_0\ak}.
\end{eqnarray}

\noindent
Using definition \jed {nenormirana_stanja}  of $\ketn\gamma$ and performing integration, we obtain

\pocjed{Runge3}
f(\a)=\sum\limits_{l=0}^\infty\sum\limits_{k=0}^{d_l} c_k^{(l)} \sum\limits_{n=0}^\infty
\sum\limits_{m=k}^{n+k} {n\choose m-k} \sqrt{{m!}\over{n!}} \;\; z_0^{n-m+k}   \;\;\ket n \bra m\;\rme^{-z_0 \ak},
\end{equation}

\noindent
which for $\alpha\in\Omega$ gives:

\pocjed{Mittag3a}
f(\a)\ket\alpha=f(\alpha)\ket\alpha .
\end{equation}

\noindent
We can rewrite \jed {Runge3}, collecting coefficients at dyads, as

\begin{eqnarray}\label{eq:Runge5}
f(\a) &=& \hat\chi\rme^{-z_0 \ak},\\
\ \ \ \hat\chi &=& \sum\limits_{l=0}^\infty\sum\limits_{n=0}^\infty\sum\limits_{m=0}^\infty\chi_{nm}^{(l)} \ket n
\bra m\ ,\nonumber\\
\chi_{nm}^{(l)} &=& \sqrt{{m!}\over{n!}} \sum\limits_{k=p}^{s_l} c_k^{(l)}{n\choose m-k} \; z_0^{n-m+k}\ ,\nonumber \\
\ \ \ p&=&\max\{0,m-n\},\;\;s_l=\min\{m,d_l\} .\nonumber
\end{eqnarray}

\subsection{Some properties of analytic functions of annihilation operator}

\noindent
Considering the basic relation between the annihilation and creation operator
$\lbrack\op{a},\op{a}^\dagger\rbrack=1$, one would expect

\begin{eqnarray}
\lbrack f(\op{a}),\op{a}\rbrack&=& 0,\label{eq:rel1}\\
\lbrack f(\op{a}),\op{a}^\dagger\rbrack&=&f^{'}(\op{a})\label{eq:rel2}
\end{eqnarray}

\noindent Since $f(\a)$ is not an entire function of $\a$, relations \jed{rel1} and \jed{rel2} need to be proven
directly. \noindent Relation \jed{rel1} can be calculated explicitly using \jed{Runge5}.
\noindent
To prove
relation \jed{rel2} we first need to construct $f^{\prime}(z)$ . It is easy to see that

\begin{eqnarray}\label{eq:izvod}
f^{\prime}(z)&=&\sum\limits_{l=0}^\infty\sum\limits_{k=0}^{d_l -1}\bar{c}_k^{(l)}(z-z_0)^k,\ z,z_0\in\Omega,\\
\ \ \bar{c}_k^{(l)}&=&(k+1)c_{k+1}^{(l)}.\nonumber
\end{eqnarray}

\noindent
Using \jed{izvod} and \jed{Runge5} we can prove \jed{rel2} directly.

\section{Logarithm of the annihilation operator}

In this section we construct an operator $\ln \a$, which is a good starting point for construction of the phase
operator \cite{arsa1},\cite{arsa}. A complex function $f(z)=\ln z$ is analytic on a simply connected subdomain
$\Omega$ in $\mathbb{C}$, which is obtained by making a cut in the complex plain along a ray originating at zero.
For simplicity, we chose a cut along the negative part of the $x$ axis. We also set $z_0=1$ in \jed{runge1}. Any
other choice of a cut and $z_0$ leads to an equivalent construction of $\ln \a$. \noindent As already noted, the
Runge theorem does not explicitly give coefficients $c_{k}^{(l)}$ in \jed{runge1}. A convenient method for
computing $c_k^{(l)}$ for the function $f(z)=\ln z$ is Mittag-Leffler expansion in the star \cite{markushevic}.
This method gives explicit constants in the expansion of $\ln z$:

\begin{eqnarray}\label{eq:ln0}
\ln z&=&\lim_{p \rightarrow \infty}\sum\limits_{l=1}^p\sum\limits_{k=1}^p\ c_k^{(l)}\ (z-1)^k\ ,\ z\in\Omega,\\
\ c_{k}^{(l)}&=&d_{k}^{(l)}\ {{(-1)^{(k+1)}}\over{k}}, k>0,\nonumber\\
\ d_{k}^{(l)} &=& {k!\over {l!}}\ \Gamma^k_p\ \Theta^l_p\ E^{(k)}_l \ ,\nonumber
\end{eqnarray}

\noindent
where

\begin{eqnarray}\label{eq:ln3}
\Gamma_p&=&2 H^{-2}_p,\\
\Theta_p &=& 1- e^{-{1\over{2}} H^2_p},\nonumber\\
E^{(k)}_l &=& {1\over{k!}}{{\partial^k}\over{\partial \rho ^k}} \prod_{s=0}^{l-1} (\rho + s)
\biggm\vert_{\rho=0}=c(l,k) .\nonumber
\end{eqnarray}
c(l,k) is unsigned Stirling number of the first kind. $H_p$ is any sequence satisfying the following condition:

\pocjed{ln6}
2 H_p^2\ e^{{1\over{2}} H^2_p} < p ,\ \ \ \ \  p>0.
\end{equation}

\noindent
This sequence of polynomials converges locally uniformly to $\ln z$ on $\Omega$. \noindent Using the
obtained coefficients in the expansion \jed{ln0} and relation \jed{Runge5} we can represent $\ln \op a$ as the
following limit:

\pocjed{ln7}
\ln\op{a} = \lim_{p \rightarrow
\infty}\sum\limits_{l=0}^p \sum\limits_{n=1}^p\sum\limits_{m=1}^p
\chi_{nm}^{(l)} \ket n \bra m \;\; \rme^{- \ak} ,
\end{equation}

\begin{displaymath}
\chi_{nm}^{(l)} = \sqrt{{m!}\over{n!}} \sum\limits_{k=k1}^m
c_k^{(l)} {n\choose m-k}        \;\;\; ,k1=\max\{0,m-n\} .
\end{displaymath}

\noindent
$\ln\op{a}$ has nice properties needed for construction of the phase operator \cite{arsa}. Using
\jed{rel1} and \jed{rel2} it is obvious that: \pocjed{ln9} \lbrack\op{a}^\dagger \op{a},-i \ln\op{a}\rbrack=
\op{a}^\dagger \lbrack\op{a},-i \ln\op{a}\rbrack + \lbrack\op{a}^\dagger,-i \ln\op{a}\rbrack\op{a}
={i\over{\op{a}}}\ \op{a} = i.
\end{equation}
We can conclude that $\ln \op{a}$ is conjugate to number operator,
and therefore is a good base for construction of the phase
operator.

\section{$\ln \op{a}$ and the phase operator}

\noindent
Let

\pocjed{faza1}
\op{\Phi}=-i \op{Y}^{-1} \ln\op{a} \op{Y} ,
\end{equation}
where

\pocjed{faza2} \op{Y}=\sum\limits_{k=0}^\infty{1\over{\sqrt{k!}}} \ket k \bra k .
\end{equation}
$\op{Y}$ is a diagonal operator so it is obvious

\pocjed{faza3} \lbrack\op{a}^\dagger \op{a}\ ,\op{\Phi}\rbrack=i.
\end{equation}
On the other hand, vectors \jed{vektor_faze_def} and coherent
states \jed{koherent1} can be combined:

\pocjed{faza4} \ket{\varphi}_f={e^{1\over{2}}\over{\sqrt{2\pi}}}
\op{Y}^{-1} \ket{e^{i\varphi}}
\end{equation}

\noindent Vector $\ket{\varphi}_f$ is an eigenvector of the
operator $\op{\Phi}$ in equation \jed{faza1}:

\pocjed{faza5} \op{\Phi} \ket{\varphi}_f = -i \op{Y}^{-1}
\ln\op{a} \op{Y} {e^{1\over{2}}\over{\sqrt{2\pi}}} \op{Y}^{-1}
\ket{e^{i\varphi}} = -i {e^{1\over{2}}\over{\sqrt{2\pi}}}
\op{Y}^{-1}  (i\varphi) \ket{e^{i\varphi}}=\varphi
\ket{\varphi}_f,
\end{equation}

\begin{displaymath}
\varphi\in(-\pi,\pi).
\end{displaymath}
Hence, the operator $\op{\Phi}$ is a phase operator.

\section{Appendix}

\noindent
To determine action of the operator $f(a)$ defined in \jed{Runge3}, on the vector $\ket\alpha$, we first
compute the following matrix element:

\pocjed{app1}
\bra{m} e^{-z_0 \op{a}^\dagger} \ket\alpha =
\norma\alpha \bra{m} \ketn {\alpha-z_0} =\norma\alpha
{{{(\alpha-z_0)^m}}\over{\sqrt{m!}}}
\end{equation}
Using \jed{Runge3} we can see

\begin{eqnarray}\label{eq:app2}
f(\a)\ket\alpha &=& \sum\limits_{l=0}^\infty\sum\limits_{k=0}^{d_l} c_k^{(l)} \sum\limits_{n=0}^\infty
\sum\limits_{m=k}^{n+k}{n\choose m-k}{1\over{\sqrt{n!}}}\;\; z_0^{n-m+k} \norma\alpha (\alpha-z_0)^m\;\;\ket n\nonumber\\
&=&\norma\alpha\sum\limits_{l=0}^\infty\sum\limits_{k=0}^{d_l} c_k^{(l)} (\alpha - z_0)^k \sum\limits_{n=0}^\infty
{1\over{\sqrt{n!}}} \sum\limits_{m=0}^{n} {n\choose m} z_0^{n-m} (\alpha - z_0)^m \;\;\ket n \nonumber
\end{eqnarray}
Last sum in previous equation is $\alpha^n$, and finally

\pocjed{app3}
f(\a)\ket\alpha=\norma\alpha\sum\limits_{l=0}^\infty\sum\limits_{k=0}^{d_l} c_k^{(l)} (\alpha - z_0)^k
\sum\limits_{n=0}^\infty {{(\alpha - z_0)^n}\over{\sqrt{n!}}} \;\;\ket n = f(\alpha)\ket\alpha
\end{equation}

\ack
The autor would like to thank Drs M. Arsenovi\'c, D. Arsenovi\'c, D. Davidovi\'c and J.Ajti\'c for their
suggestions and help in preparation of this manuscript.


\section*{References}
\begin{thebibliography}{99}
\bibitem{carruthers}
Carruthers P and Nieto M 1968 {\it Rev. Mod. Phys.}, {\bf 40} 411
\bibitem{leonhardt}
Leonhardt U 1997 {\it Measuring the quantum state of light} (Cambridge university press)
\bibitem{Pegg}
Pegg D.T. and Barnett S.M. 1988 {\it Europhys. Lett.}, {\bf 6} 483
\bibitem{levi}
Levy-Leblond J M 1976 {\it Ann. Phys. (N.Y.)} {\bf 101} 319
\bibitem{rad1_arxiv}Petrovi\' c A,  \href{http://arxiv.org/abs/1001.0777}{arXiv:1001.0777}.
\bibitem{arsa1}
Davidovi\'c Lj, Arsenovi\'c D, Davidovi\'c M and Davidovi\'c D M
2009  {\it J.Phys.A: Math. Theor.}  {\bf 42}
\bibitem{arsa}
Davidovi\'c M, Arsenovi\'c D and Davidovi\'c D M 2006  {\it J.
Phys.: Conf. Ser.}  {\bf 36} 46
\bibitem{saff}E.B. Saff, "Proceedings of Symposia in Aplied Mathematic Vol.36", (1986).
\bibitem{markushevic}Markushevich A I 1977 {\it Theory of functions of a complex variable} {\bf 2} (Chelsea)
\bibitem{alfors}Ahlfors L V 1979 {\it Complex Analysis}
\end{thebibliography}
\end{document}